\documentclass[12pt]{iopart}
\usepackage{graphicx}
\eqnobysec

\newcommand{\beq}{\begin{equation}}
\newcommand{\beqa}{\begin{eqnarray}}
\newcommand{\eeq}{\end{equation}}
\newcommand{\eeqa}{\end{eqnarray}}
\newcommand{\abs}[1]{\left\vert#1\right\vert}
\newcommand{\cas}{\smallskip\noindent $\bullet$ {\hskip 3pt}}
\newcommand{\comport}[2]{\,\mathrel{\mathop{#1}\limits_{#2}^{\null}}\,}
\renewcommand{\d}{{\rm d}}

\newcommand{\dif}[2]{{\frac{\d #1}{\d #2}}}
\newcommand{\ddif}[3]{{\frac{\d^{#3}#1}{\d #2^{#3}}}}
\renewcommand{\e}{{\rm e}}
\newcommand{\eps}{\varepsilon}
\newcommand{\frat}[2]{\ts{\frac{#1}{#2}}}
\newcommand{\ii}{{\rm i}}
\newcommand{\mean}[1]{\langle#1\rangle}
\newcommand{\ts}{\textstyle}
\newcommand{\var}{\mathop{\rm var}\nolimits}
\newcommand{\F}{{\mathrm{F}}}

\begin{document}

\title{On stochastic differential equations with random delay}

\author{P L Krapivsky$^{1,2}$, J M Luck$^2$, and K Mallick$^2$}

\address{$^1$ Department of Physics, Boston University, Boston, MA 02215, USA}

\address{$^2$ Institut de Physique Th\'eorique, IPhT, CEA Saclay
and URA 2306, CNRS, 91191 Gif-sur-Yvette cedex, France}

\begin{abstract}
We consider stochastic dynamical systems defined by differential equa\-tions
with a uniform random time delay.
The latter equations are shown to be equivalent
to deterministic higher-order differential equations:
for an $n$-th order equation with random delay,
the corresponding deterministic equation has order $n+1$.
We analyze various examples of dynamical systems of this kind,
and find a number of unusual behaviors.
For instance, for the harmonic oscillator with random delay,
the energy grows as $\exp((3/2)\,t^{2/3})$ in reduced units.
We then investigate the effect of introducing a discrete time step $\eps$.
At variance with the continuous situation,
the discrete random recursion relations thus obtained have intrinsic fluctuations.
The crossover between the fluctuating discrete problem
and the deterministic continuous one as $\eps$ goes to zero
is studied in detail on the example of a first-order linear differential equation.
\end{abstract}

\eads{\mailto{pkrapivsky@gmail.com},\mailto{jean-marc.luck@cea.fr},
\mailto{kirone.mallick@cea.fr}}

\section{Introduction}
\label{intro}

The most celebrated stochastic differential equation in Physics
is undoubtedly the Langevin equation
\beq
\dif{x(t)}{t}=\eta(t)
\label{lan}
\eeq
describing (overdamped) Brownian motion,
where $\eta(t)$ is a Gaussian white noise.
Langevin equations are usually written down
according to the following phenomeno\-logical scheme:
one starts with a deterministic equation, e.g.~$\d x/\d t=0$
in the absence of any external force,
and adds a noise term which mimics the interactions of the system
with its external environment.
The resulting Langevin equation is in general much more involved
than the original deterministic equation.
For a wide class of problems,
Langevin equations however remain
analytically tractable~\cite{Wax,risken,kampen,W,R,gardiner}.

Another broad class of dynamical systems of interest comprises memory.
Pheno\-mena involving memory effects and time delays are indeed ubiquitous.
To be more specific,
differential equations with delay play an important role
in many fields of the physical sciences, technology, economy,
and so on, whenever memory effects must be taken into account.
They have been the subject of extensive mathematical studies~\cite{delay1,delay2,delay3}.
In such a circumstance, one must know the whole past of the system,
in addition to its present, in order to predict its immediate future.
The simplest example of a differential equations with delay
is provided by the linear equation
\beq
\dif{x(t)}{t}=x(t-1),
\label{cstdelay}
\eeq
with a constant delay equal to 1.
To solve this equation,
we need to know the initial values $x(t)$ on the interval [0,1], say.
We can then recursively derive $x(t)$ at all subsequent times.

The goal of the present work is to investigate the joint effects
of stochasticity and of memory
by considering differential equations with unbounded random time delay.
Because of this memory effect, the resulting stochastic dynamics are non-Markovian.
The prototype of such an equation is the following first-order linear equation:
\beq
\dif{x(t)}{t}=x(\tau(t)),
\label{eq}
\eeq
where $\tau(t)$ is an earlier time (i.e., $0\le\tau(t)\le t$),
chosen according to some random process,
the delay itself being the time difference $t-\tau(t)$.
At variance with the Langevin equation~(\ref{lan}),
no explicit noise appears in~(\ref{eq}),
so that the random delay process $\{\tau(t)\}$ is the only source of stochasticity.
In the following we focus our attention onto the simplest
situation of a uniform sampling where, at each time $t$,
the value of $\tau(t)$ is chosen uniformly over the whole past,
i.e., in the interval $[0,t]$, independently of what occurs at other times.

Discrete avatars of the present problem,
namely linear recursions with unbounded discrete random delays,
have a long history dating back to the pioneering investigation
of the random Fibonacci sequence by Mark Kac~\cite{feigen}.
Stochastic differential equations with unbounded random delay
of the form~(\ref{eq}) have however not been investigated so far,
to the best of our knowledge.
Several mathematical works have established a range of results
on difference or differential equations with random delay~\cite{ff,kad,car,cra},
concerning the stability
and the ergodic behavior of such stochastic dynamical systems.
More specifically,
the case of differential equations with a stationary random delay is dealt
with in~\cite{car}, the emphasis being on the existence of a stationary random solution
and on its stability under discretization,
whereas a multiplicative ergodic theorem is proved
in~\cite{cra} for a class of difference equations with random delay.
The relationship between the latter investigations
and the present work is however tenuous.
It is worth emphasizing that the random delay $\tau(t)$ considered here
is {\it not} a stationary process, because it is uniformly distributed
in the growing time interval $[0,t]$.
The general results derived in~\cite{ff,kad,car,cra}
therefore do not apply in the present case.
In particular, we shall find that the typical growth
of the function $x(t)$ solving~(\ref{eq}) is a stretched exponential
of the form $\exp(2\sqrt{t})$,
whereas it is proved in~\cite{cra} that for a
stationary random delay the growth is always exponential.
Finally, yet another field of application of random delays can be found in recent
generalizations of the Black-Scholes model, where the non-uniformity
of the information in the market and the time delay until new information
becomes generally available are taken into account~\cite{shepp}.

The setup of this article is as follows.
Section~\ref{flsec} contains a detailed study
of the first-order linear equation~(\ref{eq}).
Our main result is that the solution to this equation is self-averaging.
Equation~(\ref{eq}) will indeed be shown to be equivalent
to the deterministic integro-differential equation~(\ref{eqeq}),
which boils down to the second-order differential equation~(\ref{eqdiff}).
In Section~\ref{othersec} we investigate a few other examples
of dynamical systems with random delay,
involving either higher-order time derivatives or non-linearities.
Section~\ref{discretesec} is devoted to the discrete counterpart of~(\ref{eq}),
namely the random linear recursion
\beq
x_{n+1}=x_n+\eps x_m,
\label{introeps}
\eeq
defining a random Fibonacci sequence.
The special case where $\eps=1$, originally considered by Kac~\cite{feigen},
has also been studied in~\cite{BK},
whereas the case of an arbitrary time step $\eps$ has been investigated in~\cite{KRT}.
Here we put the main emphasis on the crossover
between the fluctuating discrete problem
and the deterministic continuous one as the time step $\eps$ goes to zero.
Section~\ref{discussion} contains a brief discussion of our findings.

\section{A first-order equation with random delay}
\label{flsec}

In this section we investigate the first-order
differential equation~(\ref{eq}) in full detail.
As the latter equation is linear, it suffices to
consider the initial condition $x(0)=1$.

\subsection{Examples of deterministic delay}

In order to forge our intuition,
let us first consider a few examples
where the delay process $\tau(t)$ is deterministic.

\cas In the absence of delay, i.e., $\tau(t)=t$,
we have the ordinary differential equation
$\d x/\d t=x$, whose solution is a pure exponential:
\beq
x(t)=\e^t.
\label{noi}
\eeq

\cas With a constant delay $a$, i.e., $\tau(t)=t-a$,
we obtain an equation similar to~(\ref{cstdelay}).
The initial data must be the full function $x(t)$ for $0\le t\le a$.
The solution still grows asymptotically exponentially, albeit with a reduced rate,~as
\beq
x(t)\sim\e^{\omega t},
\eeq
where $\omega<1$ is the real solution of the characteristic equation
\beq
\omega=\e^{-\omega a}.
\label{char}
\eeq
The rate $\omega$ decreases monotonically as a function of the delay $a$.
For $a\ll1$, we have $\omega=1-a+3a^2/2+\cdots$
In the opposite regime ($a\gg1$), we have $\omega\approx(\ln a)/a$.
The characteristic equation~(\ref{char}) also has an infinite sequence of complex roots,
describing the damped oscillations displayed by $x(t)$
for generic initial data,
before the asymptotic exponential growth sets in.

\cas For an arbitrarily slowly varying delay $a(t)=t-\tau(t)$,
the solution can be argued to grow as
\beq
x(t)\sim\exp\left(\int_0^t\omega(s)\,\d s\right),
\label{atdef}
\eeq
where the instantaneous growth rate $\omega(t)$ is related to the delay $a(t)$ as
\beq
\omega(t)=\e^{-\omega(t)a(t)},
\label{atres}
\eeq
so that $\omega(t)\approx(\ln a(t))/a(t)$ is small
whenever $a(t)$ is large but slowly varying.

\cas For a delay growing linearly in time, i.e., $\tau(t)=bt$, where $b<1$ is fixed,
the solution to~(\ref{eq}) has the exact series representation
\beq
x(t)=\sum_{k\ge0}b^{k(k-1)/2}\,\frac{t^k}{k!}.
\eeq
The saddle-point method yields the asymptotic growth law
\beq
x(t)\sim\exp\left(\frac{(\ln t)^2}{2\abs{\ln b}}\right),
\eeq
which is faster than any power law, but slower than any stretched exponential.

The growth law in all these cases is vastly different,
so it is not clear what to expect in the stochastic case
of a uniform random delay.
We shall investigate the properties of equation~(\ref{eq})
by successively calculating the average of the variable $x(t)$
and its fluctuations.

\subsection{First moment}

Let us begin our analysis of~(\ref{eq}) with a uniform random delay
by studying the first moment (average) $\mean{x(t)}$.
To do so, we are led to introduce two quantities:
\beq
M(t)=\mean{x(t)},\qquad
N(t)=\int_0^t\mean{x(\tau)}\,\d\tau.
\label{mn}
\eeq
Here and throughout the following, brackets $\mean{\cdots}$ denote an average
over the realizations of the random delay process $\{\tau(t)\}$.
The above quantities obey the differential equations
\beq
\dif{M}{t}=\frac{N}{t},\qquad\dif{N}{t}=M,
\label{eqmn}
\eeq
with initial values $M(0)=1$, $N(0)=0$.

The average $M(t)$ therefore obeys the second-order differential equation
\beq
\dif{\null}{t}\left(t\dif{M}{t}\right)=t\ddif{M}{t}{2}+\dif{M}{t}=M.
\label{eqm2}
\eeq
The latter equation can be solved explicitly by looking for
the series expansion of $M(t)$.
Its solution is
\beq
M(t)=\sum_{k\ge0}\frac{t^k}{k!^2}=I_0(2\sqrt{t}),
\label{isol}
\eeq
where $I_0$ is the modified Bessel function.
We note that equation~(\ref{eqm2}) can alternatively
be transformed into a Bessel equation by using the change of variable
$T=2\sqrt{t}$.

The average solution exhibits a stretched exponential growth of the form
\beq
M(t)\approx\frac{\e^{2\sqrt{t}}}{\sqrt{4\pi}\,t^{1/4}}.
\label{iasy}
\eeq
Putting this growth law in perspective with~(\ref{atdef}) and~(\ref{atres}),
we are led to conclude that the relevant values of the time delay
lie in the range $a(t)=t-\tau(t)\sim\sqrt{t}$.

\subsection{Second moment}

In order to explore the distribution of $x(t)$,
we now turn to the second moment $\mean{x(t)^2}$.
In analogy with~(\ref{mn}), we are led to introduce three quantities:
\beqa
&&Q(t)=\mean{x(t)^2},\qquad
R(t)=\int_0^t\mean{x(\tau)x(t)}\,\d\tau,\nonumber\\
&&S(t)=\int_0^t\!\int_0^t\mean{x(\tau)x(\tau')}\,\d\tau\,\d\tau',
\label{qrs}
\eeqa
which obey
\beq
\dif{Q}{t}=\frac{2R}{t},\qquad\dif{R}{t}=Q+\frac{S}{t},\qquad\dif{S}{t}=2R,
\label{eqqrs}
\eeq
with initial values $Q(0)=1$, $R(0)=S(0)=0$.

It can easily be checked that the solution to the above equations is
\beq
Q(t)=M(t)^2,\qquad
R(t)=M(t)N(t),\qquad
S(t)=N(t)^2.
\label{qrssol}
\eeq
Indeed, as a consequence of~(\ref{eqmn}) and~(\ref{eqqrs}),
both sides of each identity of~(\ref{qrssol}) obey
the same first-order differential equations, with the same initial values.
The meaning of the above identities will be discussed in the next section.

\subsection{Deterministic behavior}
\label{detersec}

\begin{figure}[!ht]
\begin{center}
\includegraphics[angle=-90,width=.45\linewidth]{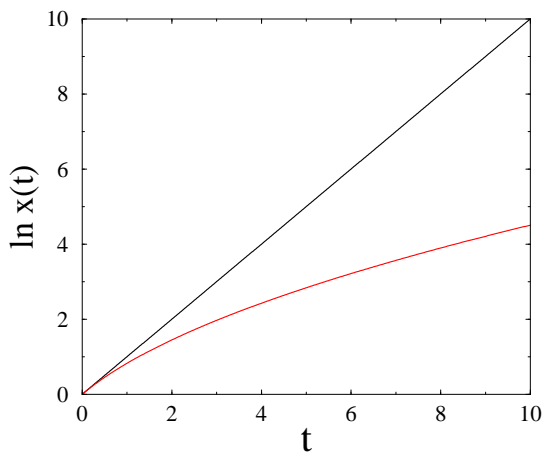}
\caption{\label{i}
Logarithmic plot of the solution $x(t)$ to the differential equation~(\ref{eq}).
Black: no delay (see~(\ref{noi})).
Red: random delay (see~(\ref{isa})).}
\end{center}
\end{figure}

The first of the identities~(\ref{qrssol}) tells us that we have identically
$\mean{x(t)^2}=\mean{x(t)}^2$ for all times $t$.
This implies that the solution to the differential equation~(\ref{eq}) with random delay
is with certainty equal to the deterministic function
\beq
x(t)=M(t)=I_0(2\sqrt{t}).
\label{isa}
\eeq
This solution is compared in Figure~\ref{i}
with the exponential solution~(\ref{noi}) in the absence of delay.

The solution~(\ref{isa})
to the differential equation~(\ref{eq}) is therefore {\it self-averaging.}
This absence of fluctuations is striking at first sight.
The stochastic differential equation~(\ref{eq}) is indeed driven
by the random delay process $\{\tau(t)\}$.

A first observation to be made is that
this self-averaging property is a peculiarity of the continuous-time limit.
The discrete counterpart of the differential equation~(\ref{eq}),
i.e., the random linear recursion~(\ref{introeps}),
indeed exhibits the generic kind of fluctuations
to be expected in a stochastic dynamical system,
including non-trivial Lyapunov exponents~\cite{feigen,BK,KRT}.
The crossover between the fluctuating discrete problem
and the deterministic continuous one
as the time step $\eps$ goes to zero
will be investigated in detail in Section~\ref{discretesec}.
More generally, applying any discretization scheme to
the stochastic differential equation~(\ref{eq})
(to be necessarily used e.g.~in a numerical analysis)
will break the self-averaging property
and resurrect statistical fluctuations.

The peculiar self-averaging property
of the continuous problem can be understood in two complementary ways.

On the physical side, the stochastic differential equation~(\ref{eq})
samples the process $\{\tau(t)\}$ for infinitely many different times,
however small $t$ is.
Fluctuations are therefore cut down by an infinite noise reduction factor,
and so the resulting process is deterministic.
In the presence of a discrete time step $\eps$,
the random delay process at time $t$ is only sampled a large but finite number of times,
of order $t/\eps$, so that the reduced variance of the process $x(t)$
is expected to be proportional to $\eps$.
This prediction will be corroborated by the analysis of Section~\ref{2sec}.

On the mathematical side,
our choice for the delay process,
namely that $\tau(t)$ is drawn at each time $t$
independently of what occurs at all the other instants,
implies that a typical realization of $\tau(t)$ is very wild function of $t$,
which is nowhere continuous, and not integrable.
The formal short-time expansion of the solution,
\beq
x(t)=1+t+\int_0^t\tau(s)\d s+\cdots,
\eeq
thus involves, as its first non-trivial term, an integral which is ill-defined.
The most natural definition of the latter integral~\cite{Svante,Michel}
just consists in replacing it by its average value, i.e.,~$t^2/4$,
since $\mean{\tau(s)}=s/2$.

An efficient way of expressing the self-averaging property
of the continuous problem is to say that~(\ref{eq}) is equivalent to
the deterministic integro-differential equation
\beq
\dif{x(t)}{t}=\frac{1}{t}\int_0^t x(\tau)\,\d\tau,
\label{eqeq}
\eeq
which boils down to the second-order differential equation
\beq
t\ddif{x}{t}{2}+\dif{x}{t}=x.
\label{eqdiff}
\eeq

To sum up, the self-averaging property
is intrinsic to the continuous-time random delay process.
It is therefore expected to hold quite generally
for dynamical systems described by differential equations with random delay,
involving either higher-order time derivatives or non-linearities.
The net effect of the random delay is to increase the order of the
differential equation by one unit.
Several examples will be dealt with in Section~\ref{othersec}.

\section{A panorama of equations with random delay}
\label{othersec}

\subsection{Oscillations induced by delay in a first-order linear equation}

We start our panorama of differential equations with random delay
by considering the equation
\beq
\dif{x(t)}{t}=-x(\tau(t)),
\label{eqj}
\eeq
which is obtained from~(\ref{eq}) by changing the sign of the rate of evolution.
This formally amounts to changing the sign of time.

In the absence of delay, we have the differential equation
$\d x/\d t=-x$, whose solution relaxes exponentially fast to zero, as
\beq
x(t)=\e^{-t}.
\label{noj}
\eeq

With random delay, $x(t)$ obeys the integro-differential equation
\beq
\dif{x(t)}{t}=-\frac{1}{t}\int_0^t x(\tau)\,\d\tau.
\eeq
It is therefore ruled by the second-order differential equation
\beq
t\ddif{x}{t}{2}+\dif{x}{t}=-x,
\eeq
whose solution is
\beq
x(t)=\sum_{k\ge0}\frac{(-t)^k}{k!^2}=J_0(2\sqrt{t})
\comport{\approx}{t\to\infty}\frac{\cos(2\sqrt{t}-\pi/4)}{\sqrt{\pi}\,t^{1/4}},
\label{jsol}
\eeq
where $J_0$ is the Bessel function.

The solution $x(t)$ exhibits a very slow power-law decay in $1/t^{1/4}$,
modulated by oscillations which slow down in the course of time.
In particular, $x(t)$ vanishes at an infinite sequence of instants
$t_k=j_k^2/4\approx(\pi^2/4)k^2$, where the $j_k$ are the zeros of $J_0$.
We have $t_1=1.445796\dots$, $t_2=7.617815\dots$, $t_3=18.721751\dots$, and so on.

\begin{figure}[!ht]
\begin{center}
\includegraphics[angle=-90,width=.45\linewidth]{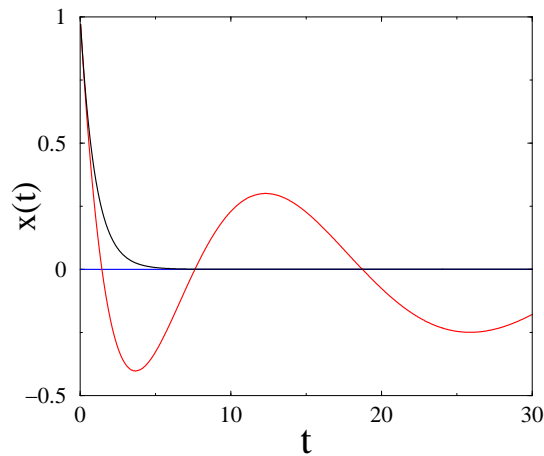}
\caption{\label{j}
Solution $x(t)$ to the differential equation~(\ref{eqj}).
Black: no delay (see~(\ref{noj})).
Red: random delay (see~(\ref{jsol})).
Blue: stable fixed point ($x=0$).}
\end{center}
\end{figure}

Figure~\ref{j} shows a comparison between the solution~(\ref{noj})
in the absence of delay, falling off exponentially fast to zero,
and~(\ref{jsol}) with random delay, relaxing in a very slow and non-monotonic~way.

\subsection{The harmonic oscillator with random delay}

The harmonic oscillator with random delay
provides another illustration of the above framework.
It is defined by the dynamical equation
\beq
\ddif{x(t)}{t}{2}=-x(\tau(t)),
\label{eqho}
\eeq
in reduced units.
We assume the initial conditions are $x(0)=1$, $\d x(0)/\d t=0$.

In the absence of delay, we have
\beq
x(t)=\cos t.
\label{noho}
\eeq
The trajectory remains bounded.
Its total energy,
\beq
E(t)=\frac{1}{2}\left(x(t)^2+\left(\dif{x}{t}\right)^2\right),
\label{hoe}
\eeq
is conserved and equals $E=1/2$.

With random delay, $x(t)$ obeys
\beq
\ddif{x(t)}{t}{2}=-\frac{1}{t}\int_0^t x(\tau)\,\d\tau,
\eeq
so that
\beq
t\ddif{x}{t}{3}+\ddif{x}{t}{2}=-x.
\eeq
The solution to this third-order differential equation is
\beq
x(t)=\sum_{k\ge0}\frac{(-2t^2)^kk!}{(2k)!^2}
={\null}_0\F_2\left(\frac{1}{2},\frac{1}{2};-\frac{t^2}{8}\right),
\label{hosol}
\eeq
where ${\null}_0\F_2$ is a hypergeometric series.

The asymptotic behavior of $x(t)$ at late times
is drastically affected by the random delay.
Setting $z=-t^2$ in the series expansion of~(\ref{hosol}),
we obtain the following estimate by means of the saddle-point method,
including its absolute prefactor:
\beq
\tilde x(z)\comport{\approx}{z\to+\infty}
\frac{1}{2\sqrt{3}}\exp\left(\frac{3}{2}\,z^{1/3}\right).
\eeq
For a positive time $t$, we have $z^{1/3}=(-t^2)^{1/3}=\e^{\pm 2\pi\ii/3}t^{2/3}$.
Summing the contributions of both complex conjugate saddle points,
we end up with the following stretched exponential growth for the position,
modulated by stretched oscillations:
\beq
x(t)\approx\frac{1}{\sqrt{3}}
\exp\left(\frac{3}{4}\,t^{2/3}\right)
\cos\left(\frac{3\sqrt{3}}{4}\,t^{2/3}\right).
\eeq

Let us keep defining the total energy of the delayed harmonic oscillator by~(\ref{hoe}).
This quantity is asymptotically dominated by its first (i.e., potential) term.
It therefore grows as $E(t)\approx x(t)^2/2$.
The kinetic energy is relatively suppressed by a factor of order~$t^{-2/3}$.
Finally, the relevant values of the time delay
are in the range $a(t)=t-\tau(t)\sim t^{1/3}$.

\begin{figure}[!ht]
\begin{center}
\includegraphics[angle=-90,width=.45\linewidth]{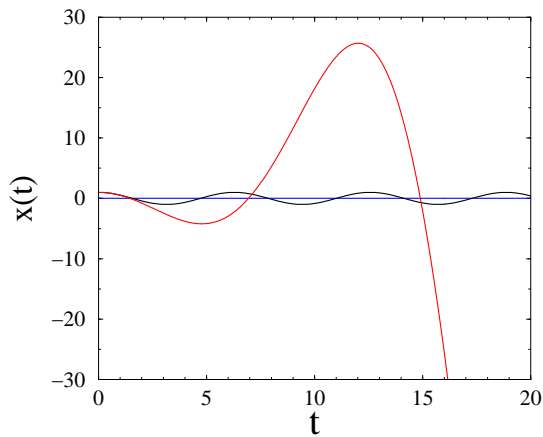}
\caption{\label{ho}
Solution $x(t)$ describing the harmonic oscillator.
Black: no delay (see~(\ref{noho})).
Red: random delay (see~(\ref{hosol})).
Blue: fixed point ($x=0$).}
\end{center}
\end{figure}

Figure~\ref{ho} shows a comparison between the periodic solution~(\ref{noho})
describing the harmonic oscillator in the absence of delay
and the growing solution~(\ref{hosol}) with random~delay.

\subsection{Higher-order linear equations}
\label{hlsec}

We now turn to the higher-order analogues of~(\ref{eq}), namely
\beq
\ddif{x(t)}{t}{m}=x(\tau(t)),
\label{eqm}
\eeq
where $m\ge1$ is an arbitrary integer.
For definiteness we assume that the initial condition is $x(0)=1$,
whereas the first $m-1$ derivatives vanish at $t=0$.

In the absence of delay, we have the differential equation $\d^m x/\d t^m=x$,
whose solution is
\beq
x(t)=\sum_{k\ge0}\frac{t^{mk}}{(mk)!}
=\frac{1}{m}\sum_{\ell=0}^{m-1}\exp(\e^{2\pi\ii\ell/m}t)
\comport{\approx}{t\to\infty}\frac{\e^t}{m}.
\eeq
This solution grows asymptotically exponentially in $t$, with unit rate,
irrespectively of the order $m$.

With random delay, $x(t)$ obeys
\beq
\ddif{x(t)}{t}{m}=\frac{1}{t}\int_0^t x(\tau)\,\d\tau,
\eeq
so that
\beq
t\ddif{x}{t}{m+1}+\ddif{x}{t}{m}=x.
\label{eqm2m}
\eeq
The solution to the latter equation is
\beqa
x(t)&=&\sum_{k\ge0}
\frac{\Gamma\!\left(\frac{1}{m}\right)}
{\Gamma\!\left(k+\frac{1}{m}\right)}
\frac{t^{mk}}{m^k(mk)!}\nonumber\\
&=&{\null}_0\F_m\left(\frac{1}{m},\frac{1}{m},\frac{2}{m},\dots,\frac{m-1}{m};
\frac{t^m}{m^{m+1}}\right),
\label{mmsol}
\eeqa
where ${\null}_0\F_m$ is a generalized hypergeometric series.
The saddle-point method yields a stretched exponential growth of the form
\beq
x(t)\sim t^{(m-2)/(2(m+1))}\exp\left(\frat{m+1}{m}\,t^{m/(m+1)}\right).
\eeq
Both the stretching index $m/(m+1)$
and the exponent $(m-2)/(2(m+1))$ of the power-law prefactor
increase with the order $m$.
The relevant values of the time delay
are in the range $a(t)=t-\tau(t)\sim t^{1/(m+1)}$.

\subsection{A non-linear first-order equation with quadratic coupling}
\label{nlsec}

In order to illustrate the effect of random delay on non-linear dynamical systems,
we focus our attention onto the case study
of a first-order equation with a quadratic non-linearity with a positive
or a negative coupling.

\subsubsection*{Positive coupling.}

In the case of a positive coupling, we consider the differential equation
\beq
\dif{x(t)}{t}=x(\tau(t))^2,
\label{q1e}
\eeq
in reduced units.
For definiteness we set again $x(0)=1$.

In the absence of delay, we obtain the equation $\d x/\d t=x^2$,
whose solution
\beq
x(t)=\frac{1}{1-t}
\label{q1sol}
\eeq
blows up in a finite time.

In the presence of a random delay,
the self-averaging property implies that $x(t)$
is again a deterministic function and that it obeys
\beq
\dif{x(t)}{t}=\frac{1}{t}\int_0^t x(\tau)^2\,\d\tau,
\eeq
so that
\beq
t\ddif{x}{t}{2}+\dif{x}{t}=x^2.
\label{q1}
\eeq
At variance with the previous examples of linear equations,
the non-linear differential equation~(\ref{q1}) cannot be solved by analytical means.
In qualitative analogy with~(\ref{q1sol}),
the solution to~(\ref{q1}) can however be expected to diverge in a finite time.
A local analysis indeed shows that $x(t)$ exhibits
a quadratic divergence of the form:
\beq
x(t)\comport{=}{t\to t_0}\frac{6t_0}{(t-t_0)^2}+\frac{12}{5(t-t_0)}-\frac{7}{25t_0}
+\frac{14(t-t_0)}{125t_0^2}+\cdots
\label{nlexp}
\eeq
The only unknown in the above expansion is the divergence time $t_0$,
which depends on the initial condition.
For $x(0)=1$ we obtain $t_0\approx3.140857$.
The effect of a random delay is therefore that the solution blows up later, but faster
(see Figure~\ref{q}, left).

\subsubsection*{Negative coupling.}

The case of negative coupling, i.e.,
\beq
\dif{x(t)}{t}=-x(\tau(t))^2,
\label{q2e}
\eeq
is formally obtained from the previous one by changing the sign of time.
For definiteness we set again $x(0)=1$.

In the absence of delay, the solution
\beq
x(t)=\frac{1}{1+t}
\label{q2sol}
\eeq
falls off slowly to zero.

In the presence of a random delay, $x(t)$ obeys
\beq
t\ddif{x}{t}{2}+\dif{x}{t}=-x^2.
\label{q2}
\eeq
The solution $x(t)$ decreases monotonically from $x(0)=1$,
crosses zero at a finite time $t_1\approx2.133528$,
and keeps decreasing until it diverges to $-\infty$
at a later finite time $t_0\approx17.00447$.
The behavior of $x(t)$ as $t\to t_0$ is still given by the expansion~(\ref{nlexp}),
up to a simultaneous sign change in $t$ and $t_0$.
The effect of random delay is more drastic in this case.
Instead of relaxing to zero, the solution overshoots
and diverges to $-\infty$ in a finite time.

\begin{figure}[!ht]
\begin{center}
\includegraphics[angle=-90,width=.45\linewidth]{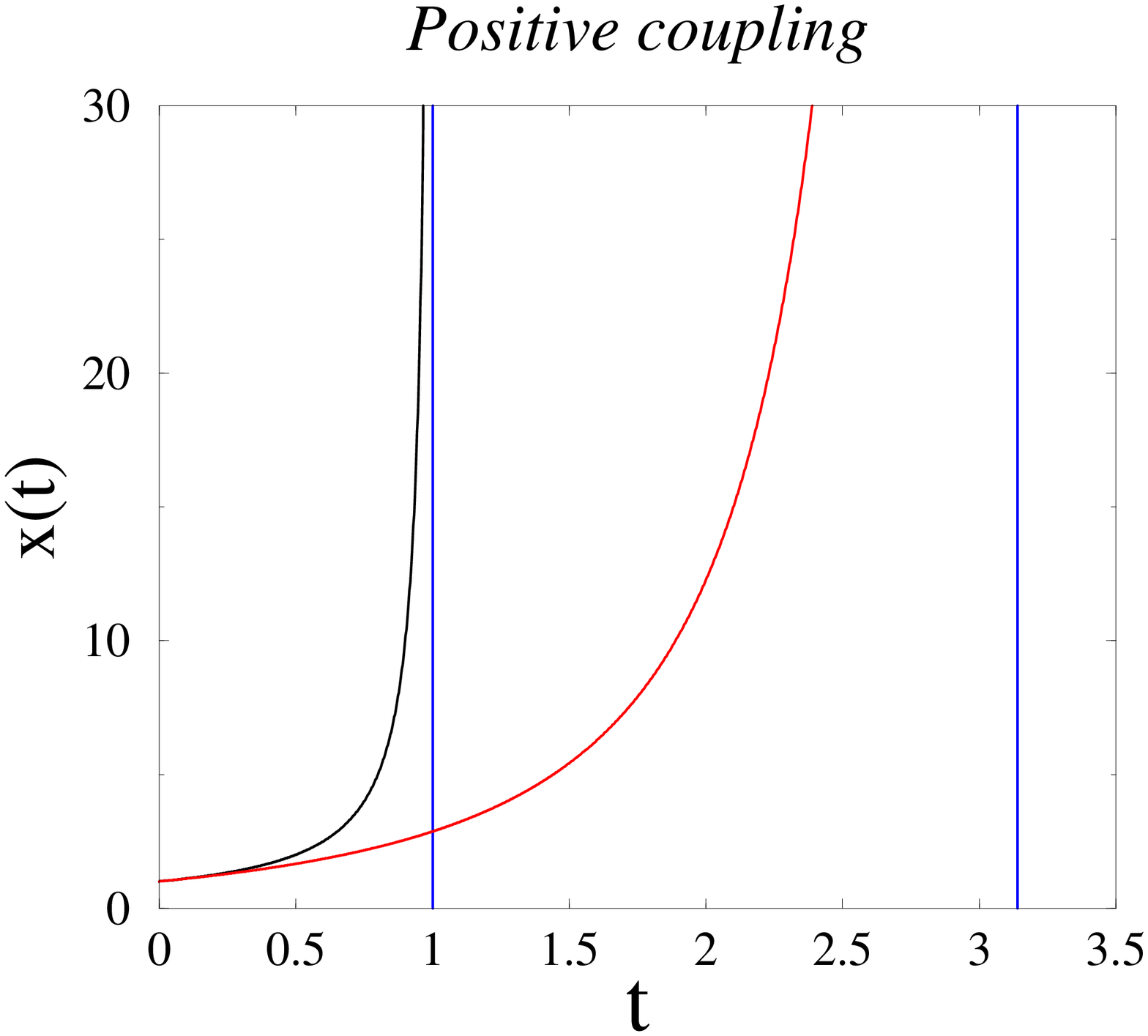}
{\hskip 10pt}
\includegraphics[angle=-90,width=.45\linewidth]{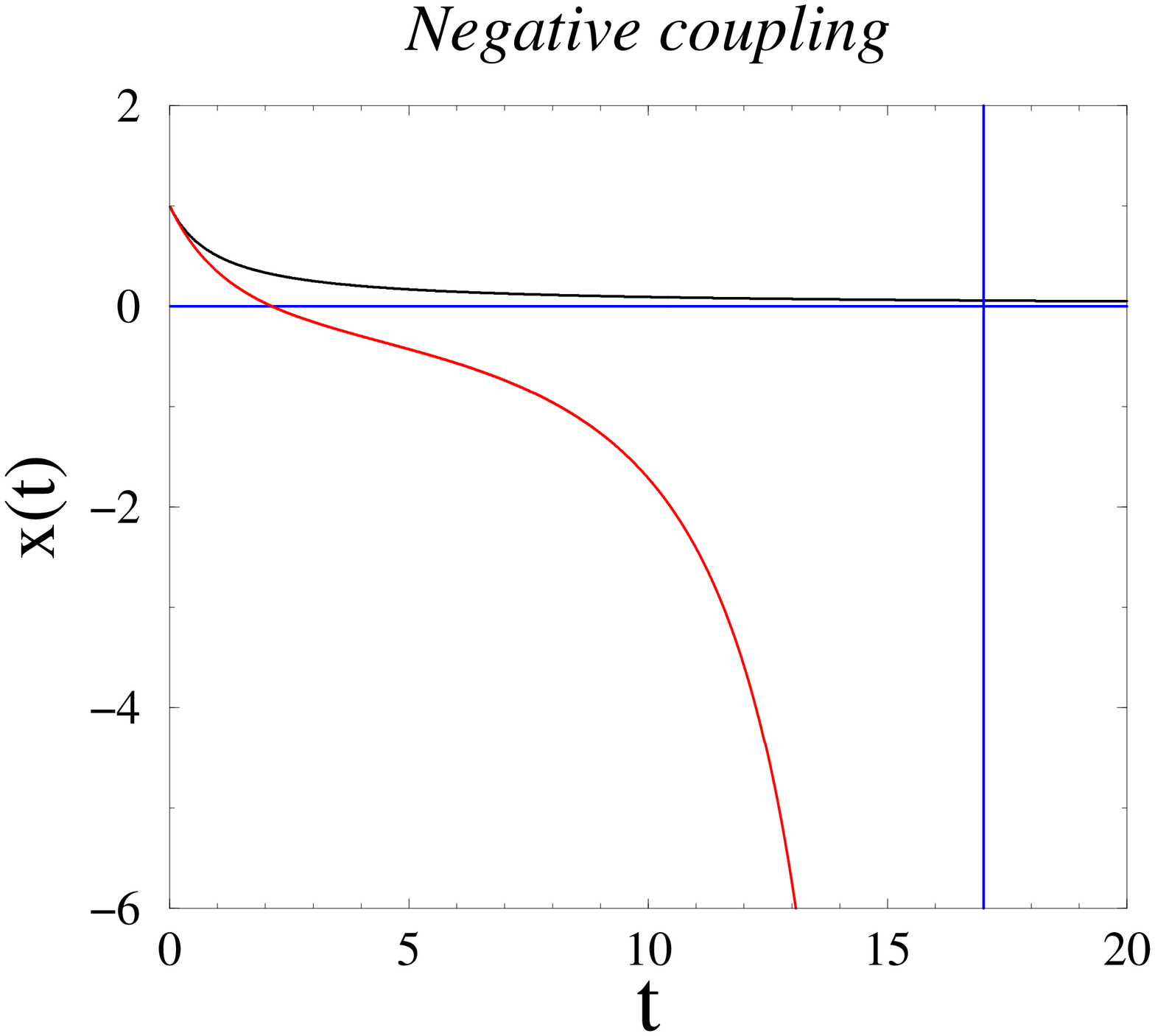}
\caption{\label{q}
Solution of the first-order differential equation with quadratic coupling.
Left: positive coupling (see~(\ref{q1e})).
Right: negative coupling (see~(\ref{q2e})).
Black: no delay (see~(\ref{q1sol}),~(\ref{q2sol})).
Red: random delay (see~(\ref{q1}),~(\ref{q2})).
Blue: asymptotes.}
\end{center}
\end{figure}

Figure~\ref{q} shows a comparison between the solutions
in the absence of delay and with random delay,
for a positive (left) and a negative (right) quadratic coupling.

\section{The discrete problem of a random linear recursion}
\label{discretesec}

This section is devoted to a detailed study of the crossover
between the fluctuating discrete problem
and the deterministic continuous one as the scale of discretization
goes to zero.
Introducing a discrete time step $\eps$ in~(\ref{eq}),
we obtain the random linear recursion (see~(\ref{introeps}))
\beq
x_{n+1}=x_n+\eps x_m,
\label{eqeps}
\eeq
leading to a random Fibonacci sequence.
At every discrete time $n$ (such that $t=n\eps$),
the label $m$ is drawn at random, uniformly among the $n+1$ integers $0,\dots,n$.
Setting $x_0=1$ for definiteness,
we have $x_1=1+\eps$,
whereas $x_2$ takes the values $1+2\eps$ and $1+2\eps+\eps^2$
with equal probabilities, and so on.

As recalled in the Introduction,
the above problem of random Fibonacci sequences has a long history,
at least in the case $\eps=1$.
The latter was indeed investigated by Mark Kac,
following a discussion with Stan Ulam.
Kac obtained the result $\beta_2=\sqrt{2(5+\sqrt{17})}$
for the growth rate of the second moment (see~(\ref{lyap2})),
and later referred to it as
{\it a tremendous formula with a square root of 17 in it}~\cite{feigen}.
The case $\eps=1$ has been further studied in~\cite{BK},
whereas the problem with an arbitrary time step $\eps$ has been investigated in~\cite{KRT}.
Our choice of~(\ref{eqeps}) as a discretization of~(\ref{eq})
follows all these earlier works.

As already announced in Section~\ref{detersec},
there is a qualitative difference between
the continuous-time differential equation~(\ref{eq})
and the discrete equation~(\ref{eqeps}).
The continuous problem has a self-averaging and deterministic solution,
whereas the discrete one exhibits the generic kind of fluctuations
to be expected in a stochastic dynamical system.
In the following we investigate many facets of the crossover
between the fluctuating discrete problem and the deterministic continuous one.

The following three regimes are worth being considered:

\cas Regime~I:\ \ $\eps\to0$ at fixed $n$ (and so $t\to0$),

\cas Regime~II:\ \ $n\to\infty$ at fixed $\eps$ (and so $t\to\infty$),

\cas Regime~III:\ \ $\eps\to0$ and $n\to\infty$ at fixed $t=n\eps$.

Let us anticipate the following stretched exponential
growth of the moments in Regime~II:
\beq
\mean{x_n^k}\sim\e^{\beta_k\sqrt{n}},\qquad
\beta_k=2k\gamma_k\sqrt{\eps}.
\label{lyaps}
\eeq
The reduced growth rates $\gamma_k$ are referred to
as the (generalized) Lyapunov exponents.

\subsection{First moment}
\label{1sec}

In analogy with the analysis of the continuous problem performed in Section~\ref{flsec},
we start by studying the first moment $\mean{x_n}$.
To do so, we introduce the quantities
\beq
M_n=\mean{x_n},\qquad
N_n=\sum_{m=0}^n\mean{x_m},
\label{MN}
\eeq
which obey the recursions
\beq
M_{n+1}=M_n+\frac{\eps}{n+1}N_n,\qquad
N_{n+1}=M_{n+1}+N_n,
\label{MNeq}
\eeq
with $M_0=N_0=1$, hence
\beq
(n+1)M_{n+1}-(2n+1+\eps)M_n+nM_{n-1}=0.
\label{M2}
\eeq
Equations~(\ref{MN})--(\ref{M2})
are the discrete analogues of~(\ref{mn})--(\ref{eqm2}).
They can be solved recursively.
We thus obtain
\beq
M_1=1+\eps,\qquad
M_2=1+2\eps+\frac{\eps^2}{2},\qquad
M_3=1+3\eps+\frac{3\eps^2}{2}+\frac{\eps^3}{6},
\eeq
and so on, so that $M_n$ is a polynomial of degree $n$ in $\eps$.

Quantitative results can be derived in the three regimes defined above.

\cas In Regime~I,
the $M_n$ have a regular expansion in the time step $\eps$:
\beq
M_n=1+n\eps+\frac{n(n-1)\eps^2}{4}+\cdots
\label{Ires}
\eeq

\cas In Regime~II,
it is convenient to introduce the generating series
\beq
F(z)=\sum_{n\ge0}M_nz^n.
\eeq
Equation~(\ref{M2}) is equivalent to the differential equation
\beq
(1-z)^2\dif{F}{z}=(1-z+\eps)F,
\eeq
whose properly normalized solution is
\beq
F(z)=\frac{\e^{\eps z/(1-z)}}{1-z}.
\label{f}
\eeq
This expression can be recognized as being the generating series
of the Laguerre polynomials~\cite{htf}.
We thus recover the result $M_n=L_n(-\eps)$~\cite{KRT}.

The asymptotic behavior of the contour integral representation
\beq
M_n=\oint\frac{\d z}{2\pi\ii}\frac{\e^{\eps z/(1-z)}}{z^{n+1}(1-z)}
\label{mint}
\eeq
can be estimated by means of the saddle-point method.
We thus obtain
\beq
M_n\approx\frac{\e^{-\eps/2}}{\sqrt{4\pi}}\,\frac{\e^{2\sqrt{n\eps}}}{(n\eps)^{1/4}}.
\label{m1}
\eeq
This expression coincides with the asymptotic behavior of the solution $M(t)$
of the continuous problem (see~(\ref{iasy})),
up to an absolute multiplicative prefactor $\e^{-\eps/2}$.
With the notation of~(\ref{lyaps}), the Lyapunov exponent of order one therefore
reads identically
\beq
\gamma_1=1.
\label{lyap1}
\eeq

The leading stretched exponential behavior of~(\ref{m1})
can be recovered without solving the difference equation~(\ref{M2}) exactly.
A more direct route can indeed be taken
if a stretched exponential behavior of the form $M_n\approx A_n\,\e^{\beta_1\sqrt{n}}$
is assumed, where $ A_n$ stands for a slowly varying prefactor.
Equations~(\ref{MNeq}) imply that a consistent hypothesis
is $N_n\approx B_n\sqrt{n}\,\e^{\beta_1\sqrt{n}}$.
Taking the continuum limit of~(\ref{MNeq}), we obtain
\beq
\frac{\beta_1}{2} A_n\approx\eps B_n,\qquad\frac{\beta_1}{2} B_n\approx A_n.
\eeq
As a consequence, $\beta_1/2$ is the eigenvalue of the $2\times2$ matrix
\beq
\Sigma_1=\pmatrix{0&\eps\cr1&0}
\eeq
whose real part is the largest.
We thus recover $\beta_1=2\sqrt{\eps}$.
This line of thought will be used for higher moments,
where an exact solution in terms of generating series will not be available.

\cas In Regime~III,
the continuum limit of the difference equation~(\ref{M2})
yields the differential equation~(\ref{eqm2}),
so that the $M_n$ converge to the solution $M(t)$ (see~(\ref{isol})).
This solution matches the behavior~(\ref{Ires}) in Regime~I,
as $M(t)=1+t+t^2/4+\cdots$ as $t\to0$.
The behavior of $M(t)$ as $t\to\infty$ also nicely matches~(\ref{m1}),
as already noticed.

\subsection{Second moment}
\label{2sec}

In order to study the second moment $\mean{x_n^2}$,
we have to introduce the four quantities:
\beqa
&&Q_n=\mean{x_n^2},\qquad
R_n=\sum_{m=0}^n\mean{x_mx_n},\nonumber\\
&&S_n=\sum_{l,m=0}^n\mean{x_lx_m},\qquad
T_n=\sum_{m=0}^n\mean{x_m^2}.
\label{QRST}
\eeqa
The variance of $x_n$ therefore reads
\beq
V_n=\var{x_n}=Q_n-M_n^2.
\eeq
The above quantities obey the recursions
\beqa
&&Q_{n+1}=Q_n+\frac{2\eps}{n+1}R_n+\frac{\eps^2}{n+1}T_n,\qquad
R_{n+1}=Q_{n+1}+R_n+\frac{\eps}{n+1}S_n,\nonumber\\
&&S_{n+1}=-Q_{n+1}+2R_{n+1}+S_n,
{\hskip 30pt}T_{n+1}=Q_{n+1}+T_n,
\label{QRSTeq}
\eeqa
with $Q_0=R_0=S_0=T_0=1$.

Four quantities are needed to evaluate the second moment in the discrete problem,
while three were sufficient in the continuous one (see~(\ref{qrs})).
The case of higher moments will be dealt with in Section~\ref{highsec}.

The above equations can be solved recursively.
We thus obtain
\beqa
&&Q_1=1+2\eps+\eps^2,\qquad
Q_2=1+4\eps+5\eps^2+2\eps^3+\frac{\eps^4}{2},\nonumber\\
&&Q_3=1+6\eps+12\eps^2+\frac{28\eps^3}{3}+\frac{25\eps^4}{6}+\eps^5+\frac{\eps^6}{6},
\eeqa
so that
\beq
V_1=0,\qquad
V_2=\frac{\eps^4}{4},\qquad
V_3=\frac{11\eps^4}{12}+\frac{\eps^5}{2}+\frac{5\eps^6}{36},
\eeq
and so on.
In general $Q_n$ and $V_n$ are polynomials of degree $2n$ in $\eps$.

Many quantitative results, most of which are novel,
can be derived in the three regimes defined above.

\cas In Regime~I,
it can be shown that $V_n$ behaves as $\eps^4$.
Skipping details, let us give the result
\beq
V_n=\frac{n(n-1)(2n+5)\eps^4}{72}+\cdots
\label{varres}
\eeq

\cas In Regime~II,
the growth of the second moment can be studied by means of the approach
sketched at the end of Section~\ref{1sec}.
Assuming a stretched exponential behavior
of the form $Q_n\approx C_n\,\e^{\beta_2\sqrt{n}}$,
we are left after some algebra with the condition
that $\beta_2/2$ is an eigenvalue of the $4\times4$ matrix
\beq
\Sigma_2=\pmatrix{0&2\eps&0&\eps^2\cr 1&0&\eps&0\cr 0&2&0&0\cr 1&0&0&0}.
\eeq
The combination $z=\beta_2^2/(4\eps)=4\gamma_2^2$ obeys the quadratic equation
\beq
z^2-(4+\eps)z+2\eps=0,
\label{quadra}
\eeq
so that the Lyapunov exponent of order two is
\beq
\gamma_2=\frac{\beta_2}{4\sqrt{\eps}}
=\frac{1}{4}\sqrt{2\left(4+\eps+\ts{\sqrt{16+\eps^2}}\right)}\;.
\label{lyap2}
\eeq
This result was first obtained in the case $\eps=1$ by Kac~\cite{feigen},
and then for an arbitrary $\eps$ in~\cite{KRT}.
For small $\eps$, the Lyapunov exponent admits the expansion
\beq
\gamma_2=1+\frac{\eps}{16}+\frac{3\eps^2}{512}-\frac{3\eps^3}{8\,192}+\cdots
\label{lyap2exp}
\eeq
The reduced second moment therefore scales as
\beq
\frac{\mean{x_n^2}}{\mean{x_n}^2}
\sim\e^{4(\gamma_2-\gamma_1)\sqrt{n\eps}}
\sim\e^{\sqrt{n\eps^3}/4}
\label{IIres}
\eeq
for $\eps\ll1$ in Regime~II
(the $\eps\to0$ limit is taken after the $n\to\infty$ limit).

\cas In Regime~III,
by taking the continuum limit of the recursions~(\ref{QRSTeq}),
we predict the following scaling behavior for the variance of the process:
\beq
V_n\approx\eps W(t).
\eeq

The variance therefore grows linearly with $\eps$, for all values of $t=n\eps$.
This result is in agreement with the argument on the noise reduction factor
exposed in Section~\ref{detersec}.
The reduced variance scales as
\beq
\frac{\var x_n}{\mean{x_n}^2}=\frac{V_n}{M_n^2}\approx\eps X(t),\qquad
X(t)=\frac{W(t)}{M(t)^2},
\eeq
where $M(t)$ is given in~(\ref{isol}).

The scaling function $W(t)$ is found to obey the third-order differential equation
\beq
t^2\ddif{W}{t}{3}+3t\ddif{W}{t}{2}+(1\!-\!4t)\dif{W}{t}-2W
=t^2\ddif{\phi}{t}{2}+3t\dif{\phi}{t}+(1\!-\!2t)\phi,
\label{eqW}
\eeq
where
\beqa
\phi(t)
&=&\frac{1}{t}\int_0^tM(s)^2\,\d s
-\left(\frac{1}{t}\int_0^tM(s)\,\d s\right)^2\nonumber\\
&=&\sum_{k\ge2}\frac{(2k)!\,t^k}{(k-2)!k!(k+1)!(k+2)!}.
\eeqa
The differential equation~(\ref{eqW}) can be interpreted as follows:
the function $\phi(t)$,
measuring the variance of the temporal dispersion of the mean solution
$M(s)$ {\it up to} time~$t$,
acts as a source for the noise variance $W(t)$ of the process {\it at} time $t$.

The solution to~(\ref{eqW}) can be written as an explicit power series.
Skipping algebraic details, we give the result
\beq
W(t)=\sum_{k\ge3}w_k\frac{(2k)!\,t^k}{k!^4},
\eeq
with
\beq
w_k=\frac{k(k^2+9k-4)}{4(k+1)(2k-1)}+\frac{3}{8}(H_{k-1}-2H_{2k-1}),
\eeq
where $H_n=\sum_{m=1}^n1/m$ is the $n$-th harmonic number.

For small times, we obtain the expansions
\beq
W(t)=\frac{t^3}{36}+\frac{t^4}{72}+\frac{59t^5}{18\,000}+\cdots,\quad
X(t)=\frac{t^3}{36}-\frac{t^4}{24}+\frac{809t^5}{18\,000}+\cdots
\eeq
The leading term matches the behavior $V_n\approx n^3\eps^4/36$
in Regime~I (see~(\ref{varres})).

In the opposite regime of late times, the saddle-point method yields
\beq
X(t)=\frac{\sqrt{t}}{4}-\frac{3\ln t}{16}+\cdots
\eeq
The reduced variance therefore grows as
\beq
\frac{\var x_n}{\mean{x_n}^2}\approx\frac{\eps\sqrt{t}}{4}\approx\frac{\sqrt{n\eps^3}}{4}.
\eeq
This power-law growth crosses over to the stretched exponential one~(\ref{IIres})
when the reduced variance becomes of order unity.
This takes place at a late time, of order $t\sim1/\eps^2$, i.e., $n\sim1/\eps^3$.

\begin{figure}[!ht]
\begin{center}
\includegraphics[angle=-90,width=.45\linewidth]{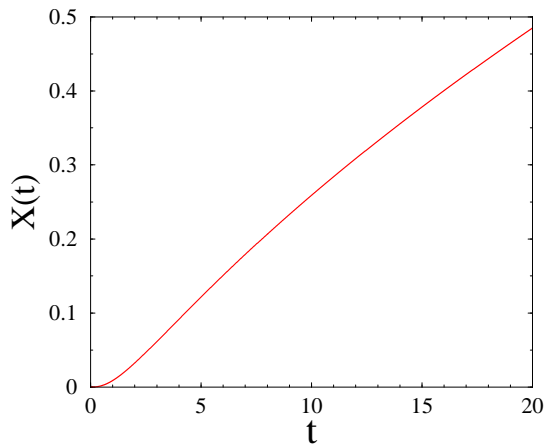}
\caption{\label{asy}
The scaling function $X(t)$ describing the reduced variance of the process.}
\end{center}
\end{figure}

Figure~\ref{asy} shows a plot of the scaling function $X(t)$
describing the reduced variance of the process.

\subsection{Higher moments}
\label{highsec}

Let us now turn to the higher moments $\mean{x_n^k}$,
where $k$ is an arbitrary integer order.
The approach exposed in the previous sections generalizes as follows.
We are led to introduce a certain number of quantities which,
together with $\mean{x_n^k}$, obey closed linear recursion relations
generalizing~(\ref{MNeq}) and~(\ref{QRSTeq}).
The number of those quantities can be evaluated as follows.
Each of them involves the product of $x_n^j$,
for some $j=0,\dots,k$, and of $k-j$ factors $x_{m_i}$,
where the labels $m_i$ are not necessarily distinct.
Attributing these labels amounts to partitioning the integer $k-j$.
For fixed $k$ and $j$, the number of distinct quantities is therefore $p_{k-j}$,
where $p_k$ is the number of partitions of the integer~$k$~\cite{htf}.
Hence the total number of quantities to be considered at order $k$~is
\beq
q_k=\sum_{j=0}^kp_j.
\eeq

Focusing our attention onto Regime~II,
and looking for a stretched exponential growth
of the form $\mean{x_n^k}\sim\e^{\beta_k\sqrt{n}}$ (see~(\ref{lyaps})),
we are left with the condition
that $\beta_k/2$ is an eigenvalue of a square matrix $\Sigma_k$ of size $q_k\times q_k$.
The eigenvalues of $\Sigma_k$ occur in pairs of opposite numbers.
The intuitive reason for that is that the coefficients of the recursion relations
are rational in $n$,
so that only even powers of $\sqrt{n}$, i.e., even powers of $\beta_k$, should matter.
If $q_k$ is even, there are $\Delta_k=q_k/2$ pairs of non-zero eigenvalues.
If $q_k$ is odd, there are $\Delta_k=(q_k-1)/2$ pairs of non-zero eigenvalues,
whereas 0 is a simple eigenvalue.
As a consequence, the combination $z=\beta_k^2/(4\eps)=(k\gamma_k)^2$
obeys a polynomial equation of degree $\Delta_k$, generalizing~(\ref{quadra}).
Table~\ref{combi} gives the numbers $p_k$,~$q_k$, and~$\Delta_k$
for the first few values of the order $k$.

\begin{table}
\begin{center}
\begin{tabular}{|c|c|c|c|c|}
\hline
$k$ & $p_k$ & $q_k$ & $\Delta_k$ & 0\\
\hline
1&1&2&1&\\
2&2&4&2&\\
3&3&7&3&$\star$\\
4&5&12&6&\\
5&7&19&9&$\star$\\
6&11&30&15&\\
7&15&45&22&$\star$\\
8&22&67&33&$\star$\\
9&30&97&48&$\star$\\
\hline
\end{tabular}
\end{center}
\caption{Numbers involved in the analysis
of the moment $\mean{x_n^k}$ up to $k=9$:
$p_k$ is the number of partitions of the integer $k$,
$q_k$ is the number of quantities needed to obtain closed recursion relations,
$\Delta_k$ is the degree of the polynomial equation
obeyed by the combination $z=\beta_k^2/(4\eps)=(k\gamma_k)^2$.
A star in the last column signals that the matrix~$\Sigma_k$ has a zero eigenvalue.}
\label{combi}
\end{table}

Let us now give some explicit results.

\cas
For $k=3$, $\Delta_3=3$, and the polynomial equation
for $z=\beta_3^2/(4\eps)=9\gamma_3^2$ is
\beqa
z^3&-&(11+4\eps+\eps^2)z^2+(19+18\eps+11\eps^2+\eps^3)z\nonumber\\
&-&(9-18\eps+16\eps^2+3\eps^3)=0.
\eeqa

\cas
For $k=4$, $\Delta_4=6$, and the polynomial equation
for $z=\beta_4^2/(4\eps)=16\gamma_4^2$ is
\beqa
z^6
&-&(25+12\eps+5\eps^2+\eps^3)z^5\nonumber\\
&+&(168+168\eps+121\eps^2+54\eps^3+10\eps^4+\eps^5)z^4\nonumber\\
&-&(400+492\eps+487\eps^2+474\eps^3+174\eps^4+37\eps^5+3\eps^6)z^3\nonumber\\
&+&(256+1104\eps-212\eps^2+712\eps^3+556\eps^4+251\eps^5+44\eps^6+2\eps^7)z^2\nonumber\\
&-&2\eps(384-224\eps-92\eps^3+124\eps^4+55\eps^5+4\eps^6)z\nonumber\\
&+&24\eps^4(2-\eps)^2=0.
\eeqa

For $\eps=1$, besides $\beta_2=\sqrt{2(5+\sqrt{17})}=4.271558\dots$,
we obtain $\beta_3=6.922223\dots$ and $\beta_4=10.120583\dots$

For small $\eps$, the Lyapunov exponents admit the expansion (see~(\ref{lyap2exp}))
\beq
\gamma_3=1+\frac{\eps}{8}+\frac{59\eps^2}{2\,304}+\frac{131\eps^3}{36\,864}+\cdots,\quad
\gamma_4=1+\frac{3\eps}{16}+\frac{91\eps^2}{1\,536}+\frac{413\eps^3}{24\,576}+\cdots
\eeq
This behavior is fully analogous to what is observed in a broad class
of disordered systems,
prototypes of which are noisy dynamical systems~\cite{Pikovsky}
and the Anderson localization problem
in one dimension~\cite{Pastur,JML,loc}.
The time step $\eps$ plays the role of the strength of disorder,
measured e.g.~by the variance of the site energies
in the case of the Anderson model with diagonal disorder.
In analogy with the latter situation,
we are tempted to deduce from the above results
that the expansion of the Lyapunov exponent of arbitrary order~$k$
(not necessarily an integer) involves polynomials in $k$ of increasing degrees,
i.e.,
\beq
\gamma_k=1+\frac{(k-1)\eps}{16}+\frac{(k-1)(32k-37)\eps^2}{4\,608}+\cdots
\label{dev}
\eeq

In the opposite regime of large $\eps$,
our explicit results for $k=2$, 3, and 4 suggest the scaling behavior
\beq
\gamma_k\approx\frac{\eps^{(k-1)/2}}{k}.
\label{large}
\eeq

\subsection{Typical sequence and fundamental Lyapunov exponent}

The scaling form~(\ref{lyaps}) of the moments
implies that the mean logarithm of $x_n$ grows as
\beq
\mean{\ln x_n}\approx2\gamma\sqrt{n\eps},
\label{log}
\eeq
where $\gamma\equiv\gamma_0$ is the fundamental (usual) Lyapunov exponent.
The latter describes the asymptotic growth law
of the most probable or typical sequence:
\beq
(x_n)_{\rm typ}\sim\exp(2\gamma\sqrt{n\eps}).
\eeq
The exact calculation of the fundamental Lyapunov exponent $\gamma$
is beyond the reach of the present work.
We can however write down its expansion at small $\eps$
by setting $k=0$ in our conjectured formula~(\ref{dev}).
We thus obtain
\beq
\gamma=1-\frac{\eps}{16}+\frac{37\eps^2}{4\,608}+\cdots
\label{dev0}
\eeq

\begin{figure}[!ht]
\begin{center}
\includegraphics[angle=-90,width=.45\linewidth]{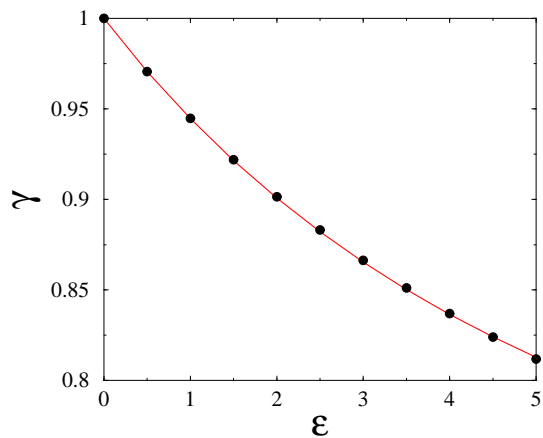}
\caption{\label{lyap}
Numerical data for the Lyapunov exponent $\gamma$ against the time step $\eps$.
Red line: rational fit incorporating the three terms of the expansion~(\ref{dev0})
and the falloff as $\eps^{-1/2}$ suggested by~(\ref{large}).}
\end{center}
\end{figure}

Figure~\ref{lyap} shows a plot of numerical data for the Lyapunov exponent $\gamma$,
obtained by means of a direct simulation of the random recursion~(\ref{eqeps}).
For each value of $\eps$,
$\ln x_n$ has been averaged over $10^7$ different realizations of $10^4$ steps each,
and the outcome fitted to~(\ref{log}).
For $\eps=1$ our estimate $\gamma\approx0.9448$, hence $2\gamma\approx1.8896$,
fully corroborates the value 1.889 given in~\cite{BK}.
The data are in very good agreement
with a fit incorporating the three terms of~(\ref{dev0}),
as well as the falloff as $\eps^{-1/2}$ suggested by~(\ref{large}).

\section{Discussion}
\label{discussion}

In this work
we have investigated the joint effects of memory (i.e., time delay)
and of extrinsic stochasticity (i.e., randomness)
on the example of differential equations with unbounded random time delay.
Our initial motivation was to investigate
continuous-time analogues of the random Fibonacci sequences
first considered by Kac~\cite{feigen}.

Stochastic differential equations with random time delay
of this kind display a self-averaging property
which leads to an unexpected deterministic behavior.
A physical explanation of this striking self-averaging behavior,
described in Section~\ref{detersec},
relies on the fact that continuous time is infinitely divisible.
So, during any time interval, whatever small,
the random delay process $\{\tau(t)\}$ will assume all
possible values because of infinite sampling,
and so the delay term will contribute only via its mean value.
In other words, the averaging method used in classical mechanics~\cite{Percival}
is exact for the present problem.
A more mathematically inclined reader
would argue that an equation with random delay
of the type~(\ref{eq}) is ill-defined because it identifies
the derivative of a function (on the left-hand side) that always satisfies some smoothness
properties (such as the intermediate value theorem)
to a wild random function (on the right-hand side)
with no smoothness property whatsoever;
therefore, the random term needs to be
regularized by replacing it by its almost-sure average value.

In any case, randomly-delayed analogues of classical dynamical systems such as
the first-order linear equation, the harmonic oscillator, or the non-linear growth model,
have been shown to behave very differently from their deterministic counterparts.

Our motivation to study continuous analogues
of random Fibonacci sequences was to calculate the Lyapunov exponent that
characterizes their typical growth by using differential methods
(which are usually more versatile and powerful than discrete techniques).
Fluctuations are however lost when taking the formal limit of a continuous time,
so that a finite time step $\eps$ has to be kept in order to preserve stochastic effects.
We have investigated the various facets of the crossover
between the fluctuating discrete problem and the deterministic continuous one.
This led us, among many other outcomes,
to the expansion~(\ref{dev}) for the generalized Lyapunov exponents
that measure the growth of the $k$-th moment.
The $k\to0$ limit of the latter result provides us with the expansion~(\ref{dev0})
for the fundamental Lyapunov exponent.
Nevertheless, an exact calculation of the latter quantity
for random Fibonacci sequences still remains a challenging open problem.

To close, it is worth mentioning that
the potentially spectacular effects of long-ranged memory on random walks
have been scrutinized in a long series of recent works~\cite{Hod,gunter,PE,CSV,K,T,KHL}.

\ack

It is a pleasure to thank Michel Bauer and Svante Janson
for illuminating discussions.


\section*{References}

\begin{thebibliography}{99}

\bibitem{Wax}
Wax N, 1954 {\it Selected Papers on Noise and Stochastic Processes} (New-York: Dover)

\bibitem{risken}
Risken H, 1984 {\it The Fokker-Planck Equation: Methods of Solution and Applications} (Berlin: Springer)

\bibitem{kampen}
van Kampen N G, 1992 {\it Stochastic Processes in Physics and Chemistry} (Amsterdam: North-Holland)

\bibitem{W}
Weiss G H, 1994 {\it Aspects and Applications of the Random Walk} (Amsterdam: North-Holland)

\bibitem{R}
Rudnick J and Gaspari G, 2004 {\it Elements of the Random Walk: An Introduction for Advanced Students and Researchers} (Cambridge: Cambridge University Press)

\bibitem{gardiner}
Gardiner C W, 2004 {\it Handbook of Stochastic Methods for Physics, Chemistry, and Natural Sciences} Springer Series in Synergetics (Berlin: Springer)

\bibitem{delay1}
Driver R D, 1977 {\it Ordinary and delay differential equations} Applied Mathematical Sciences vol 20 (New York: Springer)

\bibitem{delay2}
Lakshmikantham V, Wen L, and Zhang B G, 1994 {\it Theory of differential equations with unbounded delay} Mathematics and its Applications (Dordrecht: Kluwer)

\bibitem{delay3}
Diekmann O, van Gils S A, Verduyn Lunel S M, and Walther H O, 1995 {\it Delay equations: Functional, Complex, and Nonlinear Analysis} Applied Mathematical Sciences vol 110 (New York: Springer)

\bibitem{feigen}
Feigenbaum M, 1985 {\it An Interview with Stan Ulam and Mark Kac} J. Stat. Phys. {\bf 39} 455

\bibitem{ff}
Falin G and Fricker C, 1991 J. Applied Probab. {\bf 28} 446

\bibitem{kad}
Kadiev R I, 2004 Differential Equations {\bf 40} 276
\nonum Kadiev R I and Ponosov A V, 2007 Differential Equations {\bf 43} 898

\bibitem{car}
Caraballo T, Kloeden P E, and Real J, 2006 J. Dynamics and Differential Equations {\bf 18} 863

\bibitem{cra}
Crauel H, Doan T S, and Siegmund S, 2009 J. Difference Equations and Applications {\bf 15} 627

\bibitem{shepp}
Shepp L, 2002 IEEE Trans. Inf. Theory {\bf 48} 1372

\bibitem{BK}
Ben-Naim E and Krapivsky P L, 2002 J. Phys. A {\bf 35} L557

\bibitem{KRT}
Krasikov I, Rodgers G J, and Tripp C E, 2004 J. Phys. A {\bf 37} 2365

\bibitem{Svante}
Janson S, 2010 private communication

\bibitem{Michel}
Bauer M, 2010 private communication

\bibitem{htf}
Erd\'elyi A (ed), 1953 {\it Higher Transcendental Functions (The Bateman Manuscript Project)} (New York: McGraw-Hill)

\bibitem{Pikovsky}
Zillmer R and Pikovsky A, 2003 Phys. Rev. E {\bf 67} 061117

\bibitem{Pastur}
Lifshitz I M, Gredeskul S A, and Pastur L A, 1988 {\it Introduction to the Theory of Disordered Systems} (New-York: Wiley)

\bibitem{JML}
Luck J M, 1992 {\it Syst\`emes d\'esordonn\'es unidimensionnels} in French (Saclay: Collection Al\'ea)

\bibitem{loc}
Pendry J B, 1994 Adv. Phys. {\bf 43} 461

\bibitem{Percival}
Percival I and Richards D, 1982 {\it Introduction to Dynamics} (Cambridge: Cambridge University Press)

\bibitem{Hod}
Hod S and Keshet U, 2004 Phys. Rev. E {\bf 70} 015104
\nonum Keshet U and Hod S, 2005 Phys. Rev. E {\bf 72} 046144

\bibitem{gunter}
Sch\"utz G M and Trimper S, 2004 Phys. Rev. E {\bf 70} 045101

\bibitem{PE}
Paraan F N C and Esguerra J P, 2006 Phys. Rev. E {\bf 74} 032101

\bibitem{CSV}
da Silva M A A, Cressoni J C, and Viswanathan G M, 2006 Physica A {\bf 364} 70
\nonum Cressoni J C, da Silva M A A, and Viswanathan G M, 2007 Phys. Rev. Lett. {\bf 98} 070603
\nonum Felisberto M L, Passos F S, Ferreira A S, da Silva M A A, Cressoni J C, and Viswanathan G M, 2009 Eur. Phys. J. B {\bf 72} 427

\bibitem{K}
Kenkre V M, 2007 {\it Analytic Formulation, Exact Solutions, and Generalizations of the Elephant and the Alzheimer Random Walks} preprint arXiv:0708.0034

\bibitem{T}
Turban L, 2010 J. Phys. A {\bf 43} 285006

\bibitem{KHL}
Kumar N, Harbola U, and Lindenberg K, 2010 Phys. Rev. E {\bf 82} 021101

\end{thebibliography}
\end{document}